# C-PASS-PC: A Cloud-driven Prototype of Multi-Center Proactive Surveillance System for Prostate Cancer

Haibin Wang
Samuel Oschin Comprehensive Cancer Institute, Los Angeles, CA 90048, USA
Email: haibin.wang@cshs.org

**Abstract**

Currently there are many clinical trials using paper case report forms as the primary data collection tool. Cloud Computing platforms provide big potential for increasing efficiency through a web-based data collection interface, especially for large-scale multi-center trials. Traditionally, clinical and biological data for multi-center trials are stored in one dedicated, centralized database system running at a data coordinating center (DCC). This paper presents C-PASS-PC, a cloud-driven prototype of multi-center proactive surveillance system for prostate cancer. The prototype is developed in PHP, JQuery and CSS with an Oracle backend in a local Web server and database server and deployed on Google App Engine (GAE) and Google Cloud SQL-MySQL. The deploying process is fast and easy to follow. The C-PASS-PC prototype can be accessed through an SSL-enabled web browser. Our approach proves the concept that cloud computing platforms such as GAE is a suitable and flexible solution in the near future for multi-center clinical trials.

## 1. Introduction

Web-based data management systems offer great potential for facilitating the conduct of large scale or multi-center clinical studies [1, 2, 3, 4, 5, 6]. Investigators and researchers working across multiple sites with varying infrastructure can access data and analytical tools in these systems on a real-time basis, minimizing the logistical challenges in multi-center collaboration, providing improved monitoring capability, and facilitating new mechanisms for producing high quality validated data [3]. Traditionally, web-based data management systems for multi-center clinical studies are set up at data centers of data coordinating centers (DCCs). There is significant initial investment in infrastructure such as servers, storage, database management systems (DBMSs) and software. For example, the price tag of a dedicated database server easily exceeds $50,000. This number does not even take into account the cost of maintenance and personnel.

Cloud Computing is the long-held dream of computing as a utility, which will make software more attractive by changing it to be a service and reshape the way how IT hardware is designed and purchased [7]. It consists of the existing concepts of distributed computing [8], parallel computing [9], cluster computing [10], and utility computing [11]. Also, it is intricately related to the newly established grid computing paradigm [12]. Clinical information is the commodity used to help make patient care decisions [13]. The secure, reliable, and efficient management of clinical information is one of the key factors for the successful treatment of patients by clinicians [14]. A clinical information system is a computerized system containing an electronic patient record. It is specially designed to support users by providing complete and accurate data, and to remind practitioners by sending them reminders and alerts. Clinical information systems can also provide support for making clinical decisions, links to related medical knowledge as well as other aids [15].

In this paper we present the design and implementation of C-PASS-PC, a Cloud Computing platform based prototype of a multi-center proactive surveillance system for prostate cancer. Specifically, we exploit Google App Engine (GAE) and Google Cloud SQL-MySQL as our platform. There are other public Cloud Computing platforms available such as Amazon Web Services [16]) and Microsoft's Window Azure [17].

With this methodology a DCC can drastically reduce the cost of investment and expedite the internal review board (IRB) approval process among participating centers.

In following sections, we first introduce about Grid Computing and Cloud Computing. Then, we give a brief introduction about the clinical information system; then, we give a introduction about the background on the active surveillance of prostate cancer; then, we present the workflow and system design of proactive surveillance system for prostate cancer (PASS-PC) and deployment of the PASS-PC on GAE and Google Cloud SQL; Finally, we conclude the paper and point out the future work.

## 2. Grid Computing

Grid computing [18] enables the sharing, selection, and aggregation of a wide variety of geographically distributed resources including supercomputers, storage system, data sources, and specialized devices owned by different organizations for solving large-scale resource intensive problems in science, engineering, and commerce [19]. The motivation of Grid computing was initially driven by large-scale, resource (computational and data)-intensive scientific applications that required more resources than what a single computer (personal computer –PC, workstation, or supercomputer) could have provided in a single administrative domain.

Cancer Biomedical Informatics Grid (caBIG) [20] is a virtual network of interconnected data, services, individuals, and organizations which redefines how research is conducted, care is provided, and patients/participants interact with the biomedical research enterprise. The goals of caBIG are to: (1) adapt or build tools for collecting, analyzing, integrating, and disseminating information associated with cancer research and care; (2) connect the cancer research community through a shareable, interoperable electronic infrastructure; (3) deploy and extend standard rules. caBIG is a common language to more easily share information.

## 3. Cloud Computing

In this section, we introduce the background of Cloud Computing. We review in detail one of leading Cloud Computing platforms - GAE and Google Cloud SQL-MySQL.

Currently, almost all IT professionals talk about "Cloud Computing" as a new term for the long-held dreaming of computing as a utility, but there is no consensus on the definition of "The Cloud" [21]. In its broadest usage, "the term Cloud Computing refers to the delivery of scalable IT resources over the

Internet, as opposed to hosting and operating those resource locally" [22] . Foster et al. [12] defined Cloud Computing as "A large-scale distributed computing paradigm that is driven by economies of scale, in which a pool of abstracted, virtualized, dynamically-scalable, managed computing power, storage, platforms, and services are delivered on demand to external customers over the Internet". According to Buyya [23], "A cloud is a type of parallel and distributed system consisting of a collection of interconnected and virtualized computers that are dynamically provisioned and presented as one or more unified computing resources based on service-level agreement established through negotiation between service provider and consumers". Armbrust et al. [7] defined Cloud Computing as "The sum of Software as a Service (SaaS) and utility computing, but does not include small or medium-sized data centers, even if these rely on virtualization for management".

From these definitions of Cloud Computing by different researchers, we know that Cloud Computing is a convergence of virtualization, service-oriented architecture (SOA), distributed computing, parallel computing, and utility computing.

There are many benefits associated with Cloud Computing such as lowered costs, ability to shift capital expenses to operating expenses, agility, dynamic scalability, simplified maintenance, large scale prototyping/load testing, diverse platform support, faster management approval, and faster development [24].

### 3.1 Google App Engine and Cloud SQL

Google App Engine (GAE) is a platform for developing and hosting Web applications in Google-managed data centers. It is a Cloud Computing technology. It virtualizes applications across multiple servers and data centers. Currently, its supported programming languages are Java and Python. It also supports JRuby, JavaScript, and Scala.

GAE datastore is designed to scale, allowing apps to maintain high performance as they receive increasing traffic. Because all queries on GAE are served by pre-built indexes, the types of queries that can be executed are more restrictive than those allowed on a relational database with SQL. No joins are supported in the datastore. The datastore also does not allow inequality filtering on multiple properties or filtering of data based on results of a sub-query [25].

Because the limitations of datastore, many legacy relational database systems cannot be deployed on the GAE without a great deal of modifications to the underlying data structure.

Fortunately, Google recently announces the support of the popular open-source relational database management system MySQL on its Cloud Computing platform, which is called Cloud SQL. Currently Google Cloud SQL is available to a limited number of users.

It is well-known that Google has a very reliable and excellent performance infrastructure. GAE is built on this infrastructure. The security of GAE applications has the same policies as Google's other applications.

### 3.2 Quercus

Currently GAE only supports Java and Python programming languages. Many web applications are developed in PHP, and so these applications cannot be directly deployed on GAE.

Quercus is Caucho Technology's fast, open-source, 100% Java implementation of the PHP language [26]. Components needed to run PHP on GAE include: Java Development Kit (1.6), Google App Engine Software Development Kit (1.5.5), and Quercus & Resin Java-PHP5 bridge (4.0.18). The deployment procedure of PHP on GAE is provided in [27].

**3.3. Amazon Web Services**

The AWS are a collection of remote computing services (also called Web Services) which together constitute a Cloud Computing platform and are offered over the Internet by Amazon.com. Their most central and well-known services are Amazon Elastic Compute Cloud (Amazon EC2) and Amazon Simle Storage Service (Amazon S3).

Amazon has completed a Statement on Auditing Standards No. 70 (SAS 70) Type II Audit. It strictly controls staff's access to customer's confidential data. AWS meets the Health Insurance Portability and Accountability Act of 1996 (HIPAA) compliant requirement for transmitting individual's protected health information (PHI) across the Internet.

Although AWS has high reliability on its network, infrastructure, platform and software, it still experienced several outages in the past. For example, Amazon EC2 and S3 experienced an outage of about 3 hours on February 15, 2008; Amazon EC2 went down for about an hour for some customers in the U.S. on April 7, 2008; Amazon S3 wnt offline for about 7 hours on July 20, 2008.

Amazon EC2 Application Programming Interface (API) is becoming the de-facto API for IaaS clouds and many people are adopting it. Therefore, AWS supports horizontal interoperability among IaaS clouds which utilize Amazon EC2 APIs. For vertical interoperability, currently it is difficult to write an application which run on Amazon EC2, Window Azure, and Google App Engine.

**3.4. Microsoft Azure**

Microsoft's Windows Azure is a platform for running Windows applications and storing data in the cloud. Windows Azure runs on machine in Microsoft data centers. Instead of providing software which Microsoft customers can install and run on their local computers, Windows Azure is a Platform as a Service (PaaS) for customers to run applications and store data on Internet-accessible machines owned by Microsoft.

Windows Azure also sets up the strictest security measurements to protect customers' applications and confidential data. The Microsoft cloud undergoes annual audits for PCI DSS, SOX, and HIPAA compliance. It has obtained ISO/IEC 27001: 2005 certification and SAS 70 TYPE I and II attestations.

Microsoft utilizes a unique server balancing act which enables users to switch to another server backup when one service becomes unavailable. Fabric Controller technology reroutes work instantaneously if a server goes down, resulting in 99.9% - 99.5% uptime. But Windows Azure still experienced outage. For example, it went down for a period of 22-hours from March 13 to 14, 2009, preventing users from utilizing the eraly test release's applications.

Windows Azure allows developers to use multiple languages (.NET, PHP, Ruby, Python or Java) and development tools (Visual Studio or Eclipse) to build applications which run on Windows Azure and/or consume any of he Windows Azure platform offerings from any other cloud. Using its standard-based and interoperable approach, the Windows Azure platform supports multiple Internet protocols including HTTP, XML, SOAP, and REST key pillars of data portability.

## 4. Clinical Information System

A clinical information system (CIS) consisting of electronic patient's record is used to support the management of care. The requirements of a successful CIS include, but not limited to, the development of outcome measures, the implementation of clinical protocols and practice guidelines, and the development and implementation of decision support system such as alerts and reminders [28].

Security plays an important role in a CIS. An information security policy defines who may access what information. The term "access" includes activities such as reading, writing, appending, and deleting data. Security is driven by a threat model and it in turn improves the more detailed aspects of system design. To be effective, security policy needs to be written at the right level of abstraction without encumbering the reader with unnecessary details of specific equipment [29, 30].

Reliability is another important factor in a CIS consisting of peripherals, network, hardware, and software [31]. For example, the consequences of unexpected and unplanned downtime CIS become disastrous when these systems fail [32, 33, 34, 35]. Uptime of mission-critical clinical applications is an important marker for those who depend on that data to make decisions as well as those who monitor the operational and financial impact of systems [36].

Interoperability is another important issue of different CISs. Besides digitizing and using the information for the healthcare of their patients within organizations, clinicians, patients, and policymakers will also share appropriate information electronically between different organizations [37]. Interoperability is a fundamental requirement to ensure that widespread electronic medical record (EMR) adoption can provide our desired social and economic benefits [38].

## 5. Active Surveillance of Prostate Cancer

Prostate cancer is a cancer that forms in tissues of the prostate (a gland in the male reproductive system found below the bladder and in front of the rectum). It usually occurs in older men and is the second most common type of cancer (excluding skin cancer) among men in the United States. It is estimated that in 2011 there will be 240,890 new cases diagnosed and 33,720 deaths from prostate cancer in the United States [39]. The majority of diagnosed men will not have disease that will result in prostate cancer specific mortality; however, nearly 30,000 men will die from prostate cancer this year. The introduction of serum prostate specific antigen (PSA) screening (around 1990) led to a transient increase in prostate cancer diagnosis. Furthermore, the pattern of initial presentation of patients shifted to men with low volume disease. The long natural history of this disease has been characterized [40] and as a result has raised concerns that there may be excessive use of local intervention in men with low risk disease. This is accentutated by the recognized morbidity of the various forms of local therapy. In 2011, the initial report from the Veterans Administration sponsored PIVOT study [41] demonstrated a lock of mortality benefit for men with low-risk prostate cancer who underwent surgical intervention. Conversely, in ongoing active surveillance series, it has been shown that approximately one third of men deemed appropriate for active surveillance show evidence of progression that merits consideration of intervention while the remainder either remain stable and eventually terminate follow-up or, due to excessive anxiety, elect to proceed with therapy despite a lack of evidence of progression [42].

These data together underscore the need for understanding the natural history and biology of low-risk disease and the impact of the practice patterns of active surveillance on men with low risk disease. Several academic institutions have programs of active surveillance in which men with low-risk cancers have undergone intense observation. Low-risk cancers are typically those which 1) are small volume prostate cancers that cannot be felt on a prostate examination (digital rectal exam) and 2) lack aggressive histological morphology (microscopic appearance). These active surveillance routines have been institution specific; as such, a more comprehensive active surveillance approach is necessary. Active surveillance that is accomplished with biospecimen collection represents a key need in prostate cancer research and an evolution of this process has been coined "pro-active surveillance".

The research study entitled "Active Surveillance of Prostate Cancer" is a large-scale, multi-center clinical study sponsored by the Prostate Cancer Foundation (PCF). Cedars-Sinai Medical Center is the data coordinating center (DCC) for this study. Currently John Hopkins and Cedars-Sinai Medical Center are participating clinical sites. More clinical sites are expected to join in the study in near future.

This research study has two objectives:

1. Primary Objective

   To carefully observe men (active surveillance) with screened, detected, low-risk prostate cancer and manage them without immediate curative intervention.

2. Secondary Objective

   To explore urine and serum collected in order to develop and evaluate new and existing biomarkers for prostate cancer, evaluate biomarker changes, study gene expression profiles, and evaluate nuclear proteins.

The Proactive Surveillance System for Prostate Cancer (PASS-PC) is a Web-based, distributed, heterogeneous clinical data system developed to support this study. The system is developed by Cedars-Sinai Medical Center, The Canton Group, Grafik, John Hopkins University and the PCF. PASS-PC is HIPAA-compliant clinical data management system incorporating three main components:

1. National Prostate Surveillance Network (NPSN) website – an informational website for proactive surveillance of prostate cancer;
2. NPSN myConnect Portal – a secure patient registration web portal;
3. PASS-PC – a secure study management portal for investigators, researchers, and study coordinators.

## 6. Proactive Surveillance System for Prostate Cancer

In this section we present the workflow of PASS-PC and high-level architecture of deploying the locally-developed data system into GAE Cloud Computing platform.

Fig. 1 displays the procedure for determining patient's eligibility for the study. Fig. 2 shows the procedure of PASS-PC connectivity to NPSN and myConnect web portals.

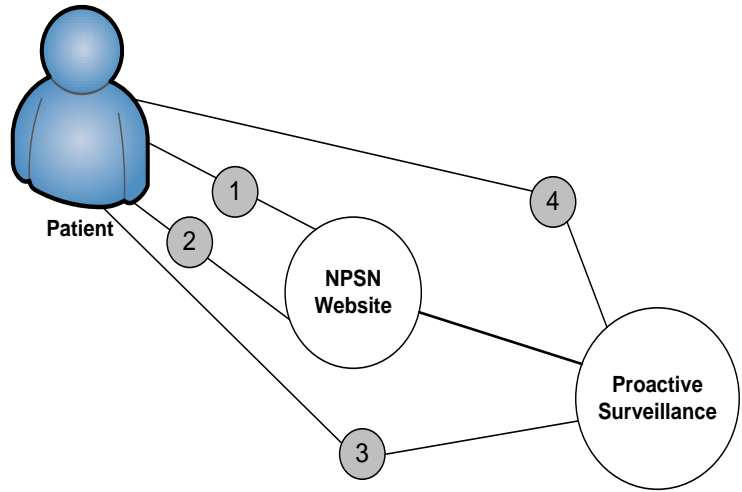

Fig. 1 The procedure for determining patient's eligibility for the study

1. Patient calculates eligibility.
2. Site returns results and next steps.
3. Patient meets with physician for initial consultation.
4. Physician validates that patient is eligible.

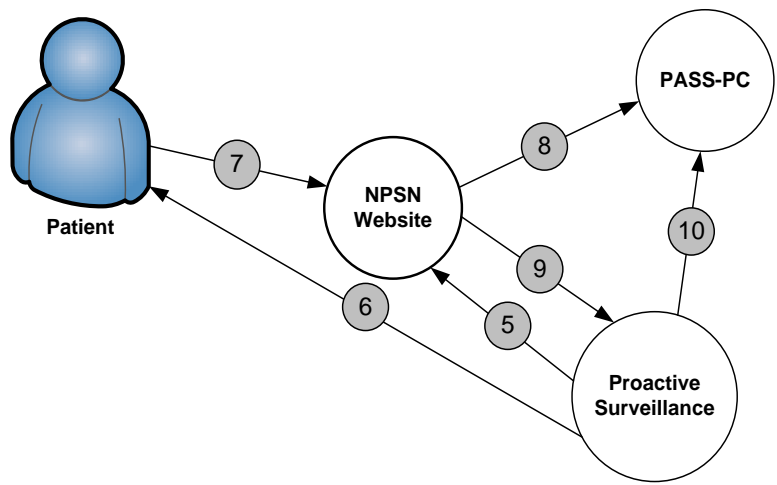

Fig. 2 The procedure of PASS-PC connectivity to NPSN and myConnect web portals

Coordinator logs into myConnect web portal and creates username and temporary password.

5. Coordinator provides username and password to patient.
6. Patient signs into myConnect web portal and completes online enrollment forms.
7. Baseline data sent to PASS-PC database when patient completed enrollment.
8. Coordinator receives auto email upon patient enrollment submission.

9. Coordinator logs into PASS-PC to access patient data (can only access own site data).

The PASS-PC high-level system architecture is displayed in Fig. 3.

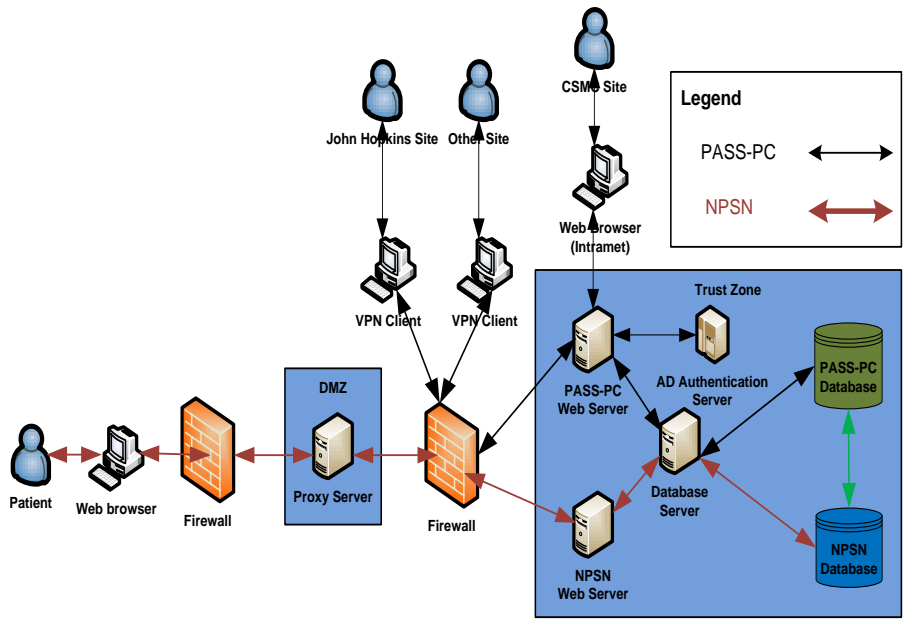

Fig. 3 The high level system architecture of the integrated PASS-PC

The PASS-PC is developed using PHP, JQuery and CSS with Oracle 11g as the backed datastore. The effort and cost of developing this system is substantial. So far the IRB application for the study protocol has not yet been approved. It is natural to consider taking advantage of a Cloud Computing platform to deploy and test the prototype of the PASS-PC, which we called C-PASS-PC, a Cloud-driven prototype of proactive surveillance system for prostate cancer. Before we deploy the PASS-PC in GAE and Google Cloud SQL, we modified the PHP code slightly to use JDBC in accessing Google Cloud SQL instead of using PDO to access Oracle 11g. Then we follow the steps in [27] to deploy the PHP, JQuery, and CSS code into GAE. The whole deployment process took less than 5 minutes.

The new system architecture of C-PASS-PC is shown in Fig 4.

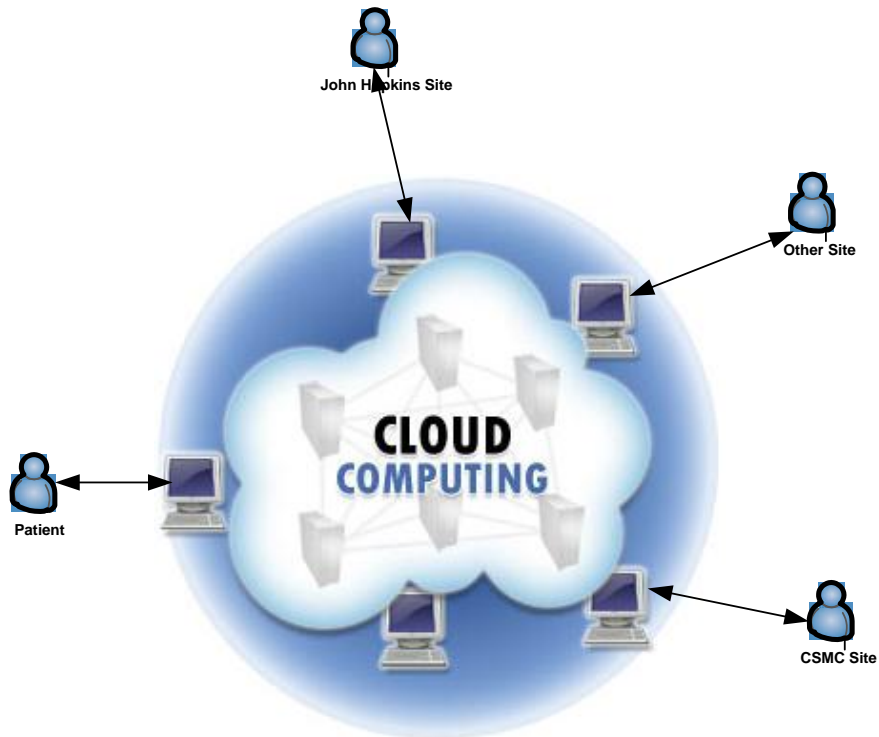

Google App Engine and Google Cloud SQL

Fig. 4 High Level System Architecture of C-PASS-PC

Comparing Fig 3.with Fig. 4 it is readily apparent that Cloud-based architecture is simple and the technical details and complexity are hidden from the system developers.

**7. Conclusion and Future work**

In this paper, we review Cloud Computing and its potential benefits. We present the design and deployment of Cloud Computing platform-based prototype of proactive surveillance system for prostate cancer (C-PASS-PC). C-PASS-PC is a Web-based multi-center clinical trial data management system. The methodology proves the concept that it is feasible and efficient to utilize Google App Engine and Google Cloud SQL to support a large-scale, multi-center clinical trial data management system. The original PASS-PC is developed by multi-part including Cedars-Sinai Medical Center, The Canton Group, Grafik, John Hopkins University and the PCF [43]. The Cedars-Sinai Medical Center data center has to set up the development infrastructure and issued vpn tokens to the outside parties to access the computing resources. With the Cloud Computing model it is easier for geographically-distributed collaborating parties to design and develop clinical trial data management.

The C-PASS-PC is just a prototype at moment. It cannot be used before the IRB application has been approved and a service level agreement has been reached among the Cloud Computing services provider and participating clinical sites.

Although GAE, Amazon AWS, and Microsoft Windows Azure can guarantee high reliability and availability, there have still been outages at Cloud Computing providers in the past. All these Cloud

Computing platforms use proprietary APIs that make interoperability more difficult. As a future work, we will study the architecture and technology of inter-Cloud communication tools which connect heterogeneous Cloud Computing platforms and present one uniform API.

**References**

1. Avidan A., Weissman C., Sprung CL., An internet web site as a data collection platform for multicenter research. Anesth Analg 2005; 100(2):506-11. PMID: 15673884.
2. Cooper C.J., Cooper S.P., del Junco D.J., Shipp E.M., Whitworth R., Cooper S.R., Web-based data collection: detailed methods of a questionnaire and data gathering tool. EpidemiolPerspectInnov 2006; 3:1. Doi: 10.1186/1742-5573-3-1.
3. Tran V-A, Johnson N., Redline S., Zhang GQ.,OnWARD: Ontology-driven web-based framework for multi-center clinical studies. J Biomed Inform , 2011, doi: 10.1016/j.jbi.2011.08.019
4. Zhang G.Q., Siegler T., Saxman P., Sandberg N., Mueller R., Johnson N., et al. VISAGE: a query interface for clinical research. In: Proceedings of the 2010 AMIA clinical research informatics summit, San Francisco, March 12-13; 2010. P. 76-80. PMCID: PMC3041531.
5. Hernandez J., Acuna C., de Castro M., Marcos E., Lopez M, Malpica N., Web-PACS for multicenter clinical trials. IEEE Trans Inf Technol Biomed, 2007; 11(1):87-93.
6. Wisniewski S., Web-based communications and management of a multi-center clinical trial: the sequenced treatment alternatives to relieve depression (STAR*D) project, Clinical Trials, 2004, doi: 10.1191/1740774504cn035oa. 1(4): 387-398.
7. Armbrust M., Fox A., Griffith R., Joseph A., Katz R., Konwinski A., Lee G., Patterson D., Rabkin A., Stoica I., Zahar M., A view of Cloud Computing. Comm. of the ACM, 2010; 53(4):50-58.
8. Sukumar G., Distributed Systems – An Algorithmic Approach, Chapman & Hall/CRC 2007.
9. Asanovic K., Dobik R., Catanzaro B., etl., The landscape of parallel computing research: a view from Berkeley, University of Berkely, Technical Report No. UCB/EECS-2006-183. 2006.
10. Bader D., Pennington R., Cluster computing: applications, The International Journal of High Performance Computing, 2001; 15(2): 181-185.
11. "On-demand computing: what are the odds?", ZD Net, Nov 2002, [accessed August 1, 2012].
12. Foster I., Zhao Y., Raicu I., Lu S., Cloud Computing and Grid Computing 360-degree compared, In: Proceedings of Grid Computing Environments Workshop, Austin, November 16; 2008. P. 1-10.
13. Wyatt J., Meidcal informatics: artefacts or science?, Methods Inf Med, 1996; 35(3):197-200.
14. Wang H.,Chen Z., Cloud Computing: Is It Ready for Clinical Information Systems*?, Proc. of JICSIT 2011*, August, 2011.
15. Dick R., Steen E., Detmer D., The computer-based patient record, National Academy Press, 1997.
16. Amazon Web Services, available at http://aws.amazon.com/. [accessed December 1, 2011].
17. Microsoft Window Azure, available at http://www.microsoft.com/windowsazure/. [accessed December 1, 2011].
18. Foster I., Kesselman C. (Eds.), The Grid: Blueprint for a Future Computing Infrastructure, Morgan Kaufmann, San Francisco, USA, 1999.
19. Buya R., Yeo C., Venugopal S., Cloud computing and emerging IT platforms: vision, hype, and reality for delivering computing as the $5^{th}$ utility, Future Generation Computer Systems, 2009; 25:599-616.

20. Eschenbach A., Buetow K., Cancer informatics vision: caBIGTM, Cancer Informatics, 2006; 2:22-24.
21. "Twenty experts define Cloud Computing", SYS-CON Media Inc, http://cloudcomputing.sys-con.com/read/612375_p.htm, 2008. accessed August 1, 2012].
22. "7 things you should know about Cloud Computing", available at http://net.educause.edu/ir/library/pdf/EST0902.pdf. [accessed August 1, 2012].
23. Buyya R., Yeo C., Venugopal S., Market-oriented Cloud Computing: vision, hype, and reality for delivering IT services as computing utilities. In: Proceedings of the 10th IEEE International Conference on High Performance Computing and Communications, Dalian, September 25-27; 2008. P. 5-13.
24. Hogan M., Cloud Computing & Databases: How databases can meet the demands of cloud computing. http://www.scaledb.com/pdfs/CloudComputingDaaS.pdf, 2008. [cited August 1, 2012].
25. "Mastering the datastore", available at http://code.google.com/appengine/articles/datastore/overview.html. [accessed December 1, 2011].
26. "Quercus: PHP in Java", available at http://www.caucho.com/resin-3.0/quercus/. [accessed August 1, 2012].
27. "PHP on Google App Engine", available at http://php-apps.appspot.com/. [accessed August 1, 2012].
28. Anderson J., "Clearing the way for physician's use of clinical information systems, Communications of the ACM, 1997; 40(8): 83-90.
29. Anderson R., Security in clinical information systems, http://www.cl.cam.ac.uk/~rjal4/Papers/policy11.pdf. [accessed August 1, 2012].
30. Anderson R., A security policy model for clinical information systems, Proc. 1996 IEEE Symposium on Security and Privacy, 1996; 30-43.
31. Littlehohns P., Wyatt M., Garvican L., Evaluating computerized health information systems: hard lessons still to be learnt, BMJ, 2003; 326(7394): 860-863.
32. Valenstein P., Treling C., Aller R., Laboratory computer availability: a college of American pathologists Q-probes study of computer downtime in 422 instituions, Arch Pahol Lab Med, 1996; 120(7): 626-632.
33. Valenstein P., Walsh M., Six-year trends in laboratory computer availability, Arch Pathol Lab Med, 2003, 127(2): 157-161.
34. Anderson M., The toll of downtime, Healthcare Informatics, 2002; 19(4): 27-30.
35. Kovach D., Developing a reliable medical informatics network, Healthcare Financial Management, 1999, 53(1): 48-49.
36. Nancy N., Downtime procedures for a clinical information system: a critical issue, J. of Critical Care, 2007; 22(1): 45-50.
37. Walker J., Pan E., Hohnston D., Adler-Milstein J., Bates D., Middleton B., The value of healthcare information exchange and interoperability, Health Affairs, January 19 2005. http://content.Helthaffairs.org/cgi/content/abstract/hlthaff.w5.10. [accessed August 1, 2012].
38. Brailer D., Interoperability: the key to the future health care system, Health Affairs, January 19 2005. http://content.healthaffairs.org/cgi/content/abstract/hlthaff.w5.19v1. [accessed August 1, 2012].
39. "Prostate Cancer", available at http://www.cancer.gov/cancertopics/types/prostate. [accessed August 1, 2012].
40. Pound C., Partin A., Eisenberger M., Chan D., Pearson J., Walsh P.. Natural history of progression after PSA elevation following radical prostatectomy. JAMA 1999; 281(17):1591-1597.

41. American Urological Association, PIVOT study. http://clinicaltrials.gov/ct2/show/NCT00007644. [accessed August 1, 2012].

42. Tosoian J., Trock B., Landis P., et al. Active surveillance program for prostate cancer: an update of the Johns Hopkins experience. J ClinOncol 2011; 29:2185-2190.

43. Wang H., Yatawara M., Huang S., Dudley K., Szekely C., Holden S., Piantadosi S., The integrated proactive surveillance system for prostate cancer, The Open Medical Informatics Journal, 2012; 6: 1-8.